\documentclass[12pt, a4paper]{article}

\usepackage[margin=1in]{geometry}
\usepackage{amsmath}
\usepackage{amssymb}
\usepackage{amsthm}
\usepackage{hyperref}
\usepackage{listings}
\usepackage{xcolor}
\usepackage{parskip}
\usepackage{microtype}
\usepackage{enumitem}
\usepackage{booktabs}
\usepackage{tabularx}
\usepackage{tikz}
\usepackage{pgfplots}
\usetikzlibrary{arrows.meta,positioning,calc}
\pgfplotsset{compat=1.18}

\hypersetup{
  colorlinks=true,
  linkcolor=blue,
  citecolor=blue,
  urlcolor=blue,
}

\title{\textbf{Behavioral Intelligence Platforms:}\\
\large From Event Streams to Autonomous Insight via\\
Probabilistic Journey Graphs, Behavioral Knowledge Extraction,\\
and Grounded Language Generation}

\author{%
  Arun Patra\thanks{Code, simulation scripts, and datasets:
    \url{https://github.com/journium/journium-research}.} \\
  \normalsize Journium, Inc. \\
  \normalsize \texttt{arun@journium.app}
  \and
  Bhushan Vadgave \\
  \normalsize Journium, Inc. \\
  \normalsize \texttt{bhushan@journium.app}
}

\date{March 2026}

\begin{document}

\maketitle

\begin{abstract}
Contemporary product analytics systems require users to pose explicit queries---writing SQL, configuring dashboards, or constructing funnels---before any insight can surface. This pull-based paradigm imposes a dual bottleneck: it demands both domain knowledge and technical fluency from practitioners who must know, in advance, which questions to ask. We argue that the next generation of behavioral analytics must invert this model, shifting from passive data stores that answer queries to active intelligence systems that continuously monitor, detect, and narrate behavioral phenomena without prompting.

We present the Behavioral Intelligence Platform (BIP), a system architecture and formal framework that transforms raw event streams into automatically generated, evidence-backed insights. BIP introduces four tightly integrated layers: (1) a Normalization and State Derivation (NSD) stage that standardizes raw events and maps them to a multi-level semantic state hierarchy; (2) a Behavioral Graph Engine (BGE) that models user journeys as absorbing Markov chains, computing transition probabilities, removal effects, and path quality metrics; (3) a Behavioral Knowledge Graph (BKG) combined with a Detector System (DS) that reifies graph outputs into a queryable triple-store of grounded behavioral facts and autonomously identifies behavioral phenomena; and (4) a Grounded Language Layer (GLL) that constrains large language model (LLM) output to verified facts from the BKG, producing faithful narrative insights at scale.

We formalize the Behavioral Intelligence Problem, define a taxonomy of detectors for autonomous insight generation (activation drivers, drop-off clusters, behavioral regressions, segment divergence, unexpected loops), and propose a composite interestingness score for prioritizing insights under bounded attention. We discuss the key design principles---grounded-first computation, semantic state abstraction, time-aware snapshots, and segment-aware analysis---and characterize the fundamental trade-offs between completeness, causal interpretability, and computational cost. We define evaluation criteria for correctness, latency, and insight quality, and characterize the fundamental challenges of assessing behavioral intelligence systems --- challenges that are unique to unsupervised, push-based insight generation and that motivate an open evaluation agenda.

\end{abstract}

\tableofcontents
\newpage

\section{Introduction}

The canonical workflow of a product analyst begins with a question. The analyst opens a dashboard, constructs a query, examines a chart, and forms a hypothesis---then iterates. This interactive, pull-based model has historically underpinned virtually every analytics platform in commercial use, from Mixpanel and Amplitude to Google Analytics and Heap. The user is assumed to be an active driver of inquiry; the system is a passive responder.

This assumption is increasingly untenable. Product teams face an ever-expanding corpus of instrumented behavior---hundreds of event types, thousands of user properties, dozens of defined funnels---against which the space of meaningful questions is combinatorially vast. Under the pull model, most of this behavioral signal goes unexamined. Insights that are not explicitly sought are not found. Critical regressions in activation flows go undetected until they manifest in lagging revenue metrics. Segment-specific friction points accumulate invisibly behind aggregate conversion numbers. Fast-moving product cycles demand a pace of inference that manual exploration cannot sustain.

We observe a structural mismatch: the volume and velocity of behavioral data have grown substantially, while the human capacity for directed inquiry has not. The solution is not more powerful query engines or more expressive visualization tools---it is a paradigm shift from pull-based to push-based intelligence, where the analytics system takes epistemic initiative.

This paper formalizes and addresses the \textit{Behavioral Intelligence Problem}: given a continuous stream of product events across a population of users or accounts, automatically detect, rank, and narrate behavioral phenomena of potential significance---without requiring the practitioner to specify what to look for.

This problem is distinct from, and more ambitious than, existing adjacent work. Automated data exploration (ADE) systems \cite{law2020,ma2021} automate the exploration of tabular datasets but lack the journey-centric, temporal, and segment-aware structure of behavioral analytics. Anomaly detection systems identify statistical deviations but do not model user journeys, cannot reason about path structure, and produce point-in-time alerts rather than behaviorally coherent narratives. Even recent benchmarks for natural language query interfaces to structured data \cite{he2024} presuppose that the user poses a question. We seek a system that generates questions and answers them simultaneously.

Our contributions are as follows:

\begin{itemize}[leftmargin=1.5em]
  \item \textbf{Problem formalization.} We formally define the Behavioral Intelligence Problem, characterizing the input (event streams), the output (a ranked feed of evidence-backed insight objects), and the objectives (high interestingness, statistical validity, actionability, and faithfulness).

  \item \textbf{Multi-level state model.} We introduce a three-level state hierarchy (raw event, semantic, lifecycle) and formalize state derivation as a rule-based mapping that enables meaningful journey abstraction while preserving traceability to raw events.

  \item \textbf{Absorbing Markov chain journey model.} We model user journeys as absorbing Markov chains over a derived state space and derive closed-form expressions for conversion probability, expected journey length, and state removal effects.

  \item \textbf{Behavioral Knowledge Graph.} We define a typed fact schema---behavioral triples of the form (subject, predicate, object) with associated evidence payloads and confidence scores---that serves as an auditable intermediate representation between numerical computation and language generation.

  \item \textbf{Detector taxonomy.} We taxonomize the class of behavioral phenomena detectable by deterministic detectors and propose an interestingness scoring framework for prioritizing the resulting insight feed.

  \item \textbf{Grounded Language Layer.} We propose an architecture for constraining LLM-generated narratives to verified knowledge graph facts, separating numerical computation from linguistic expression and systematically preventing hallucination in analytics narratives.

  \item \textbf{Design principles and trade-off analysis.} We characterize four core design principles for behavioral intelligence systems and analyze the trade-offs they entail, with reference to existing literature.
\end{itemize}

The remainder of this paper is organized as follows. Section~\ref{sec:related} reviews related work. Section~\ref{sec:problem} formally defines the Behavioral Intelligence Problem. Section~\ref{sec:architecture} describes the BIP architecture. Section~\ref{sec:formal} develops the formal foundations. Section~\ref{sec:detectors} describes the detector system. Section~\ref{sec:gll} describes the Grounded Language Layer. Section~\ref{sec:evaluation} discusses evaluation. Section~\ref{sec:discussion} analyzes limitations. Section~\ref{sec:conclusion} concludes.

\section{Related Work}
\label{sec:related}

\subsection{Automated Data Exploration and Insight Generation}

The problem of automatically generating insights from tabular data has received sustained attention in the database and visualization communities. Law et al.\ \cite{law2020} characterize the space of automated data insights, identifying properties such as novelty, deviation, and coverage as criteria for interestingness. DataShot \cite{xu2020} generates fact sheets from tabular data by enumerating a predefined set of statistical patterns (extrema, trends, comparisons) and ranking them by interestingness. MetaInsight \cite{ma2021} discovers structured knowledge from multidimensional data by identifying common patterns across subspaces via an optimal partition scheme.

More recent work has explored LLM-mediated insight generation. The QUIS system \cite{manatkar2024} uses a two-module architecture in which a question-generation module produces statistically-informed questions about a dataset and a downstream module answers them, iterating until insight quality converges. InsightBench \cite{sahu2024} provides a benchmark for evaluating end-to-end data analysis agents. InsightPilot \cite{ma2023} proposes an LLM-empowered automated data exploration system. These systems target general tabular data and do not address the journey-centric, temporal, and segment-structured nature of product behavioral data.

A key distinction between prior ADE work and the behavioral intelligence setting is the role of time and sequence. Tabular ADE systems treat records as independent observations; behavioral data is fundamentally sequential and causal, with the ordering and structure of events encoding user intent, friction, and outcome probability in ways that cross-sectional summary statistics cannot capture.

\subsection{Markov Models for User Journey Analysis}

Markov chain models have a long history in modeling sequential user behavior, including customer journey attribution across marketing touchpoints. The marketing-analytics literature has developed Markov-chain attribution models for multi-touch advertising paths: Shao and Li \cite{shao2011} introduced data-driven multi-touch attribution; Anderl et al.\ \cite{anderl2016} formalised graph-based attribution and the \emph{removal effect} of a touchpoint as a structural-importance score; Kakalejcik et al.\ \cite{kakalejcik2022} apply this framework to e-commerce. Harbich et al.\ \cite{harbich2017} propose a mixture of Markov models for customer journey map discovery, and the Customer Behaviour Hidden Markov Model \cite{wang2022} extends this to e-commerce with hidden state representations.

The absorbing Markov chain formalism provides a natural foundation for computing expected journey lengths and outcome probabilities in the presence of uncertain paths. In an absorbing Markov chain with transient states $T$ and absorbing states $A$, the fundamental matrix $N = (I - Q)^{-1}$ gives the expected number of times the chain visits each transient state before absorption, enabling closed-form derivation of conversion probabilities and expected remaining steps \cite{kemeny1960}.

BIP shares the absorbing-Markov-chain machinery and the removal-effect formulation with the marketing attribution lineage above, but differs in setting and objective: marketing attribution explains a known terminal conversion across paid touchpoints to allocate budget; BIP applies the same machinery to in-product behavioral states to autonomously \emph{detect and rank} structurally critical states across the full journey graph, integrate the resulting facts into a knowledge-graph layer, and drive grounded narrative generation. Existing journey analysis tools---Mixpanel's Flows, Amplitude's User Paths, Heap's Path Analysis---offer UI-level path visualization but do not expose the underlying probabilistic model, do not compute removal effects, and require manual interpretation.

\subsection{Process Mining}

Process mining \cite{vanderaalst2016} is a discipline concerned with discovering process models from event logs in enterprise workflows. Customer journey analysis has been studied through a process mining lens \cite{bernard2017}, leveraging CJM-structured event logs to assess journey duration and quality.

BIP is conceptually related to process mining but addresses a different setting: consumer product events rather than enterprise process logs, behavioral heterogeneity rather than conformance to a predefined process model, and automated insight generation rather than interactive exploration.

\subsection{Knowledge Graphs and Grounding}

Knowledge graphs represent entities and their relationships as typed triples (subject, predicate, object), providing a structured representation suitable for both machine reasoning and natural language generation. The integration of LLMs with knowledge graphs---under the umbrella of Knowledge Graph Retrieval-Augmented Generation (GraphRAG) \cite{li2024,zhang2025}---has emerged as a dominant paradigm for grounding language model outputs in verified external knowledge.

Existing GraphRAG systems such as SubgraphRAG \cite{li2024} retrieve relevant subgraphs and use them as context for LLM inference. These systems ground the LLM in pre-existing knowledge bases; BIP's Behavioral Knowledge Graph is dynamically constructed from live behavioral data and populated by deterministic detectors, making the grounding relationship tighter and more auditable.

Pan et al.\ \cite{pan2024} describe KGs as a `cognitive middle layer' between raw input and LLM reasoning. BIP instantiates this principle in the specific domain of behavioral analytics.

\subsection{Proactive and Augmented Analytics}

The concept of augmented analytics was identified as a key trend by Gartner \cite{gartner2019} and has since been extended toward proactive analytics paradigms \cite{gartner2025}. Commercial systems such as ThoughtSpot Spotter, Salesforce Einstein Discovery, and Microsoft Power BI Copilot offer varying degrees of automated insight generation, primarily targeting anomaly detection and natural language query interfaces.

These systems typically operate on pre-aggregated metrics and lack the journey-level behavioral modeling that characterizes BIP.

\section{Problem Formulation}
\label{sec:problem}

\subsection{Setting}

Let $E = \{e_1, e_2, \ldots, e_n\}$ denote a temporally ordered stream of events, where each event $e_i = (\text{actor\_id}, \text{event\_name}, \text{timestamp}, \text{properties}, \text{context})$ describes a discrete action taken by an actor (user or account) at a specific moment. The stream is partitioned by organization and project, and events are associated with actors whose identity may be resolved across anonymous and authenticated sessions.

A segment $S$ is a predicate over actor properties defining a cohort: $S(a) = 1$ iff actor $a$ satisfies the segment criteria. Let $J = \{S_j \mid j = 1..K\}$ be a defined set of segments. We are additionally given a set of terminal states $T$ representing outcome labels assigned to actors at the end of an observation window:
\[
T = \{\text{converted},\ \text{churned},\ \text{inactive},\ \text{retained},\ \text{dropped\_off}\}
\]

\subsection{The Behavioral Intelligence Problem}

Given a continuous event stream $E$, a set of semantic state definitions $\Phi$, journey definitions $\Delta$, segment definitions $J$, and outcome labels $T$, the Behavioral Intelligence Problem is to produce a ranked feed $I = \{(\text{insight\_type}, \text{content}, \text{confidence}, \text{evidence}, \text{timestamp})\}$ such that:

\begin{itemize}[leftmargin=1.5em]
  \item \textbf{Validity:} Every claim in an insight object can be traced to a deterministic computation over $E$ with stated statistical support.
  \item \textbf{Interestingness:} Insights are ranked by a composite score combining statistical significance, magnitude of effect, actionability, and novelty relative to prior observations.
  \item \textbf{Faithfulness:} Narrative descriptions of insights do not introduce claims not supported by the underlying evidence payload.
  \item \textbf{Coverage:} The system generates insights across the full space of detectable behavioral phenomena, not only those corresponding to user-specified queries.
  \item \textbf{Timeliness:} Insights reflect behavioral changes within a latency bound appropriate to their actionability.
\end{itemize}

\subsection{Formal Objectives and Non-Goals}

BIP does not claim to solve causal inference. The behavioral facts it produces reflect statistical associations that may or may not reflect causal relationships. The system uses association predicates (\texttt{increases\_probability\_of}, \texttt{associated\_with}, \texttt{more\_common\_in}) rather than causal predicates (\texttt{causes}, \texttt{leads\_to}) in its knowledge representation.

BIP also does not target real-time streaming at sub-second latency, nor fully learned sequence models such as transformer-based next-event predictors.

\section{System Architecture}
\label{sec:architecture}

\subsection{Overview}

\begin{figure}[htbp]
\centering
\begin{tikzpicture}[
  node distance=0.9cm,
  box/.style={rectangle, draw=black!55, fill=blue!6, rounded corners=4pt,
              text width=7cm, minimum height=1.25cm, align=center, font=\small},
  arrow/.style={-{Stealth[length=7pt,width=5pt]}, thick, black!60},
  lbl/.style={font=\footnotesize\itshape, text=black!45, midway, right=6pt}
]
  \node[font=\small\bfseries]           (input)  {Event Stream $E$};
  \node[box, below=0.8cm of input]      (nsd)    {\textbf{Layer 1: NSD}\\[3pt]
                                                   Normalization \& State Derivation};
  \node[box, below=0.9cm of nsd]        (bge)    {\textbf{Layer 2: BGE}\\[3pt]
                                                   Behavioral Graph Engine};
  \node[box, below=0.9cm of bge]        (bkg)    {\textbf{Layer 3: BKG + DS}\\[3pt]
                                                   Knowledge Graph \& Detector System};
  \node[box, below=0.9cm of bkg]        (gll)    {\textbf{Layer 4: GLL}\\[3pt]
                                                   Grounded Language Layer};
  \node[font=\small\bfseries, below=0.8cm of gll] (output) {Insight Feed $I$};

  \draw[arrow] (input)  -- (nsd);
  \draw[arrow] (nsd)    -- node[lbl]{Derived State Events}       (bge);
  \draw[arrow] (bge)    -- node[lbl]{Graph Snapshots \& Metrics} (bkg);
  \draw[arrow] (bkg)    -- node[lbl]{Facts \& Findings}          (gll);
  \draw[arrow] (gll)    -- (output);
\end{tikzpicture}
\caption{BIP four-layer data-flow architecture. Each layer produces a structured
artifact consumed by the next: NSD derives a semantic state hierarchy from raw
events; BGE builds absorbing Markov chain snapshots; BKG+DS reify graph metrics
into grounded facts and emit typed findings; GLL generates faithful narrative
insights constrained to verified facts.}
\label{fig:architecture}
\end{figure}
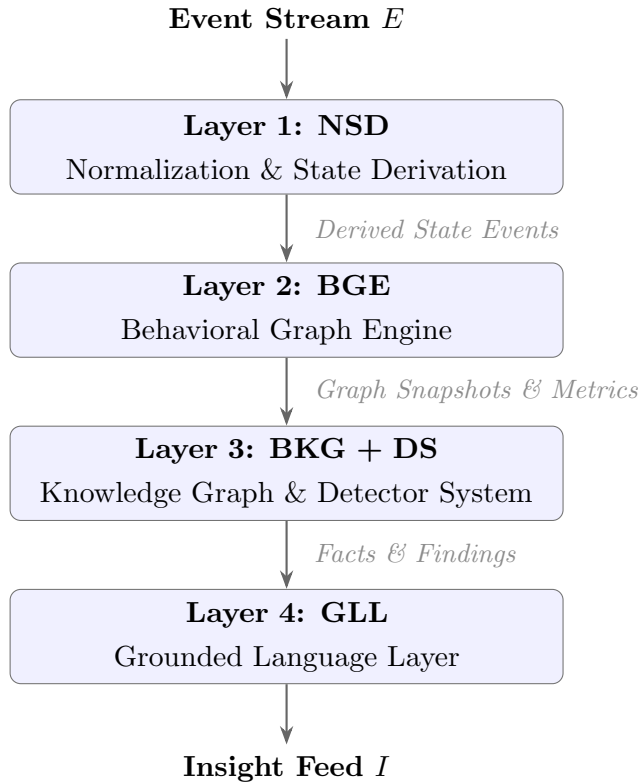

BIP is organized as a layered pipeline with four principal stages, each producing a structured artifact consumed by the next. Figure~\ref{fig:architecture} shows the high-level data flow.

\subsection{Layer 1: Normalization and State Derivation}

The normalization stage performs five operations on raw events: (i) deduplication by \texttt{event\_id} with configurable idempotency window, (ii) bot and health-check filtering via User-Agent and behavioral heuristics, (iii) identity resolution that stitches anonymous and authenticated sessions using a deterministic graph traversal over aliasing events, (iv) late arrival handling with a configurable ingestion lag tolerance, and (v) timestamp normalization to UTC with sub-millisecond precision.

Following normalization, raw events are mapped to derived state events via a three-level state hierarchy. At the \texttt{raw\_event} level, state identity is the event name itself. At the semantic level, states are defined by rule-based predicates over event properties. At the lifecycle level, states represent coarse user lifecycle positions---\texttt{new\_user}, \texttt{activated}, \texttt{paying}, \texttt{churned}, \texttt{retained}---derived from a combination of event history and time windows.

This three-level abstraction reduces state space explosion: a product with 200 raw event types typically yields 15--30 meaningful semantic states and 5--8 lifecycle states.

\subsection{Layer 2: Behavioral Graph Engine}

The BGE aggregates derived state events into a directed weighted graph $G = (V, E, W)$ where $V$ is the set of states, $E \subseteq V \times V$ is the set of observed transitions, and $W: E \to \mathbb{R}^+$ assigns transition counts. The BGE materializes the following metrics for each graph snapshot: (i) per-edge transition probabilities, (ii) per-state reach rates, outcome conversion probabilities, and expected remaining steps under the absorbing Markov chain model, (iii) top-$N$ materialized paths ranked by occurrence and conversion efficiency, and (iv) segment-conditional versions of all of the above.

\subsection{Layer 3a: Behavioral Knowledge Graph}

The BKG converts BGE outputs and detector findings into a typed triple-store of behavioral facts. Each fact takes the form:
\begin{align*}
\textit{fact} = (\, &\text{subject},\ \text{predicate},\ \text{object},\ \text{confidence}, \\
                    &\text{validity\_window},\ \text{evidence\_payload},\ \text{provenance}\,)
\end{align*}

Predicates are drawn from a closed vocabulary of behavioral relations:
\begin{itemize}[noitemsep, leftmargin=2em]
  \item \texttt{transitions\_to},\ \texttt{increases\_probability\_of},\ \texttt{is\_activation\_driver\_for}
  \item \texttt{is\_dropoff\_point\_for},\ \texttt{diverges\_from},\ \texttt{regressed\_after},\ \texttt{changed\_after}
\end{itemize}
The closed predicate vocabulary constrains the knowledge graph to a semantically well-defined ontology and ensures that retrieval patterns for the GLL remain tractable.

Each fact is bound to the snapshot from which it was derived, giving the system a complete lineage from raw events through derived states through graph metrics through facts.

\subsection{Layer 3b: Detector System}

Detectors are deterministic analysis modules that consume BGE snapshot outputs and emit typed findings and facts. The detector system runs after each snapshot build and is extensible: new detectors can be added without modifying the BGE or BKG schema.

\subsection{Layer 4: Grounded Language Layer}

The GLL translates structured findings and facts into natural language insights via a two-stage process: (i) a retrieval stage that constructs a fact bundle---the minimal set of BKG facts supporting a given finding---and (ii) a generation stage that prompts an LLM with the fact bundle and a system prompt enforcing association-not-causation language. The LLM is explicitly prohibited from inventing numerical values, computing novel statistics, or asserting causal relationships.

\section{Formal Foundations}
\label{sec:formal}

\subsection{Absorbing Markov Chain Journey Model}

We model a journey definition $\delta$ as an absorbing Markov chain $M_\delta = (\mathcal{S}, \mathcal{A}, Q, R)$ where $\mathcal{S}$ is the set of transient (non-terminal) states, $\mathcal{A}$ is the set of absorbing (terminal) states, $Q$ is the $|\mathcal{S}| \times |\mathcal{S}|$ sub-stochastic matrix of transition probabilities among transient states, and $R$ is the $|\mathcal{S}| \times |\mathcal{A}|$ matrix of transition probabilities from transient states to absorbing states.

Transition probabilities are estimated from observed journey instances. For a pair of states $(s_i, s_j)$, the empirical transition probability is:

\begin{equation}
  P(s_j \mid s_i) = \frac{\text{count}(s_i \to s_j)}{\sum_k \text{count}(s_i \to s_k)}
  \label{eq:transition}
\end{equation}

where the denominator sums over all states, including absorbing states. This satisfies the row-stochastic property: for each transient state $s_i$,

\begin{equation}
  \sum_j P(s_j \mid s_i) + \sum_t P(t \mid s_i) = 1
  \label{eq:rowstochastic}
\end{equation}

Transition probabilities are estimated over a fixed analysis window, implicitly assuming behavioral stationarity within that window. This assumption is acknowledged to be an approximation; Section~\ref{sec:discussion} discusses non-stationarity as a direction for future work.

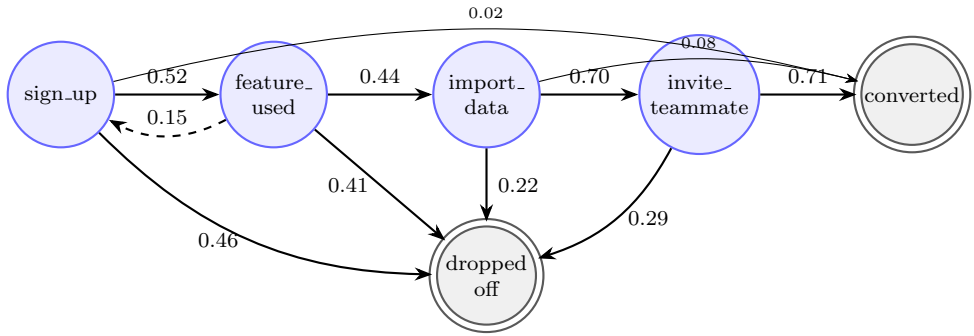
\begin{figure}[htbp]
\centering
\begin{tikzpicture}[
  transient/.style={circle, draw=blue!60, fill=blue!8,
                    minimum size=1.4cm, font=\scriptsize, align=center, inner sep=2pt},
  absorbing/.style={circle, draw=black!65, fill=gray!12,
                    minimum size=1.4cm, font=\scriptsize, align=center, inner sep=2pt,
                    double, double distance=2pt},
  >=Stealth, thick
]
  \node[transient] (s1) at (0,0)      {sign\_up};
  \node[transient] (s2) at (2.8,0)    {feature\_\\used};
  \node[transient] (s3) at (5.6,0)    {import\_\\data};
  \node[transient] (s4) at (8.4,0)    {invite\_\\teammate};
  \node[absorbing] (cv) at (11.2,0)   {converted};
  \node[absorbing] (dp) at (5.6,-2.4) {dropped\\off};

  \draw[->]        (s1) -- node[above,font=\scriptsize]{0.52} (s2);
  \draw[->]        (s2) -- node[above,font=\scriptsize]{0.44} (s3);
  \draw[->]        (s3) -- node[above,font=\scriptsize]{0.70} (s4);
  \draw[->]        (s4) -- node[above,font=\scriptsize]{0.71} (cv);

  \draw[->,dashed] (s2) to[bend left=28]
                        node[above,font=\scriptsize]{0.15} (s1);

  \draw[->,thin]   (s1) to[bend left=14]
                        node[above,font=\tiny]{0.02} (cv);
  \draw[->,thin]   (s3) to[bend left=14]
                        node[above,font=\tiny]{0.08} (cv);

  \draw[->] (s1) to[bend right=22]
                  node[left,font=\scriptsize]{0.46} (dp);
  \draw[->] (s2) -- node[left,font=\scriptsize]{0.41} (dp);
  \draw[->] (s3) -- node[right,font=\scriptsize]{0.22} (dp);
  \draw[->] (s4) to[bend left=22]
                  node[right,font=\scriptsize]{0.29} (dp);
\end{tikzpicture}
\caption{Example six-state funnel as an absorbing Markov chain.
Single-bordered circles are transient states ($\mathcal{S}$);
double-bordered circles are absorbing states ($\mathcal{A}$).
Edge labels are empirical transition probabilities; each row sums to~1
across all outgoing edges (thin arrows indicate low-probability direct
conversions). The dashed back-edge represents
re-visitation from \texttt{feature\_used} to \texttt{sign\_up}.}
\label{fig:markov}
\end{figure}

Figure~\ref{fig:markov} shows a concrete six-state funnel modeled as an absorbing Markov chain. Four transient states (\texttt{sign\_up}, \texttt{feature\_used}, \texttt{import\_data}, \texttt{invite\_teammate}) lead to two absorbing states (\texttt{converted} and \texttt{dropped\_off}). The dashed back-edge from \texttt{feature\_used} to \texttt{sign\_up} illustrates re-visitation, which is fully representable in the model.

\subsubsection{Fundamental Matrix}

The fundamental matrix

\begin{equation}
  N = (I - Q)^{-1}
  \label{eq:fundamental}
\end{equation}

provides the expected number of times the chain visits transient state $s_j$ when starting from transient state $s_i$ before absorption. The matrix $N$ exists and is finite when the chain is absorbing---a condition we enforce by design via the terminal state requirement in journey definitions.

\subsubsection{Absorption Probabilities}

The probability of being absorbed by terminal state $t$ when starting from transient state $s_i$ is given by the $(i, t)$ entry of the absorption probability matrix:

\begin{equation}
  B = N \cdot R
  \label{eq:absorption}
\end{equation}

In the product analytics context, $B(s_i, \text{converted})$ is the probability of conversion when \textbf{starting from} transient state $s_i$ --- equivalently, the long-run fraction of journeys that reach \texttt{converted} among those that begin at $s_i$.

\begin{figure}[htbp]
\centering
\begin{tikzpicture}
\begin{axis}[
  ybar,
  bar width=18pt,
  width=0.80\linewidth,
  height=6.2cm,
  ylabel={$B[s, \text{converted}]$},
  ylabel style={font=\small},
  xtick={1,2,3,4},
  xticklabels={%
    \texttt{sign\_up},
    \texttt{feature\_used},
    \texttt{import\_data},
    \texttt{invite\_teammate}},
  xticklabel style={font=\small, rotate=12, anchor=east},
  ymin=0, ymax=0.85,
  ytick={0,0.2,0.4,0.6,0.8},
  yticklabel style={font=\small},
  grid=major,
  grid style={dashed,gray!30},
  nodes near coords,
  nodes near coords style={font=\scriptsize},
  enlarge x limits=0.18,
]
\addplot[fill=blue!50, draw=blue!70] coordinates {
  (1,0.165) (2,0.279) (3,0.577) (4,0.710)
};
\end{axis}
\end{tikzpicture}
\caption{Absorption probabilities $B[s, \text{converted}]$
(Eq.~\eqref{eq:absorption}) computed by closed form from the transition
matrix in Figure~\ref{fig:markov} via $B = N \cdot R$. Each bar is the
probability that a journey starting from state $s$ ultimately reaches
\texttt{converted}. Reproduced exactly by \texttt{markov\_journey\_model.py};
the empirical conditional $P(\text{converted} \mid \text{reached}(s))$ from
Monte Carlo trajectories agrees with $B[s, \text{converted}]$ to within
sampling noise (see \texttt{simulate\_trajectories.py}), as required by the
strong Markov property when all journeys share a common starting state.}
\label{fig:absorption}
\end{figure}
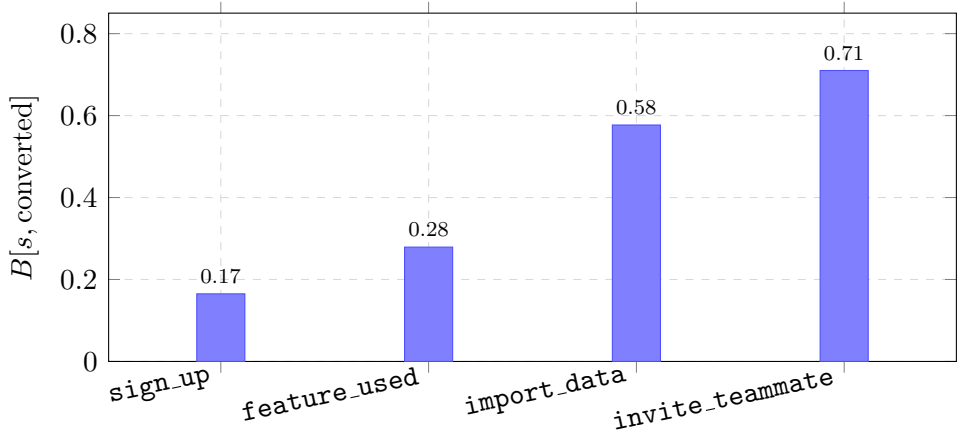

In the empirical product context we additionally track $P(\text{converted} \mid \text{reached}(s))$ --- the fraction of observed journeys that converted among all that \emph{visited} $s$ at any point. When all journeys begin at a common starting state, the strong Markov property implies $P(\text{converted} \mid \text{reached}(s)) = B(s, \text{converted})$ exactly: conditioning on reaching $s$ collapses the journey distribution to one that effectively starts at $s$, and the future evolution is conditionally independent of the past. The two measures can differ in practice when (i) trajectories are right-censored before absorption (a common artifact of rolling-window snapshots), or (ii) the starting-state distribution itself is heterogeneous and depends on the conditioning event. Figure~\ref{fig:absorption} plots $B[s, \text{converted}]$ directly; the empirical conditional from Monte Carlo trajectories matches to within sampling noise (see \texttt{simulate\_trajectories.py}).

For activation-driver detection (Section~\ref{sec:detectors}), the relevant comparison is between $B(s, \text{converted})$ and $P(\text{converted} \mid \neg\,\text{reached}(s))$ --- the conversion rate among journeys that bypass $s$ entirely. States with a large gap between these two quantities---\texttt{feature\_used}, \texttt{import\_data}, and \texttt{invite\_teammate} in this example---are strong candidate activation drivers, with import\_data the most extreme: journeys that bypass \texttt{import\_data} convert only $\sim$3\% of the time vs.\ $\sim$58\% among those that reach it (lift $\approx 20\times$).

\subsubsection{Expected Remaining Steps}

The expected number of steps until absorption from transient state $s_i$ is given by the row sum of $N$:

\begin{equation}
  t_i = \sum_j N_{ij}
  \label{eq:expectedsteps}
\end{equation}

This quantity serves as a proxy for journey friction and is used in the drop-off detector and path quality scoring.

\subsection{State Reach Rate and Conversion Lift}

The reach rate of state $s$ in a journey is the fraction of journey instances that include at least one visit to $s$:

\begin{equation}
  \text{reach\_rate}(s) = \frac{|\{j : s \in \text{states}(j)\}|}{|\text{journeys}|}
  \label{eq:reachrate}
\end{equation}

The conversion lift of state $s$ with respect to terminal outcome $t$ measures the increase in conversion probability associated with visiting $s$:

\begin{equation}
  \text{lift}(s, t) = \frac{P(t \mid \text{reached}(s))}{P(t \mid \neg\,\text{reached}(s))}
  \label{eq:lift}
\end{equation}

For states where $P(t \mid \neg\,\text{reached}(s)) = 0$, lift is undefined; the platform emits a special \texttt{necessary\_for\_conversion} finding. States with high reach rate and high lift are candidate activation drivers.

\subsection{Removal Effect}

The removal effect of state $s$ with respect to terminal outcome $t$ is the decrease in overall conversion rate when $s$ is removed from the journey graph. Removal is modeled by: (i) deleting all edges into and out of $s$ from the transition graph, (ii) re-normalizing transition probabilities from predecessor states of $s$, and (iii) recomputing the absorption probability matrix $B'$ under the modified graph:

\begin{equation}
  \text{removal\_effect}(s, t) = B(\text{start}, t) - B'(\text{start}, t)
  \label{eq:removal}
\end{equation}

A high removal effect indicates that the state lies on the dominant conversion paths in the graph. Note that removal effect is not equivalent to causal effect; it is a structural property of the observed graph.

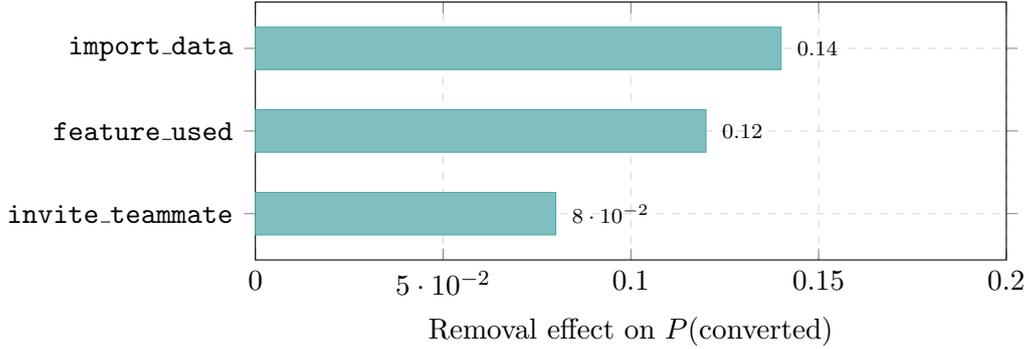
\begin{figure}[htbp]
\centering
\begin{tikzpicture}
\begin{axis}[
  xbar,
  bar width=16pt,
  width=0.72\linewidth,
  height=5.0cm,
  xlabel={Removal effect on $P(\text{converted})$},
  xlabel style={font=\small},
  ytick={1,2,3},
  yticklabels={%
    \texttt{invite\_teammate},
    \texttt{feature\_used},
    \texttt{import\_data}},
  yticklabel style={font=\small},
  xmin=0, xmax=0.20,
  xtick={0,0.05,0.10,0.15,0.20},
  xticklabel style={font=\small},
  grid=major,
  grid style={dashed,gray!30},
  nodes near coords,
  nodes near coords style={font=\scriptsize, xshift=2pt},
  enlarge y limits=0.28,
]
\addplot[fill=teal!50, draw=teal!70] coordinates {
  (0.08,1) (0.12,2) (0.14,3)
};
\end{axis}
\end{tikzpicture}
\caption{Removal effect ranking (Eq.~\eqref{eq:removal}) for the example funnel.
Each bar shows the decrease in overall $P(\text{converted})$ when the corresponding
state is removed from the journey graph and transition probabilities are
re-normalised. \texttt{import\_data} is the structurally most critical state,
as it is the sole gateway to \texttt{invite\_teammate} --- the highest-converting
transient state.}
\label{fig:removal}
\end{figure}

Figure~\ref{fig:removal} ranks non-start transient states by removal effect. \texttt{import\_data} has the highest structural impact ($\Delta \approx 0.14$): as the sole gateway to \texttt{invite\_teammate}, removing it eliminates the dominant conversion path.

The computational cost of removal effect is $O(|S|^3)$ per state per snapshot due to matrix inversion. For large state spaces, we apply it selectively to candidate states identified by a combined score of $\text{reach\_rate} \times \text{lift} \times \text{sample\_size} > \tau$.

\subsection{Temporal Change Detection}

For each edge $(s_i, s_j)$ in the behavioral graph, we compute the delta in transition probability between consecutive snapshots:

\begin{equation}
  \Delta P(s_i, s_j) = P_t(s_j \mid s_i) - P_{t-1}(s_j \mid s_i)
  \label{eq:temporal}
\end{equation}

Statistical significance of changes is assessed using a two-proportion z-test under the null hypothesis that the two proportions are equal.

For release-linked change detection, the comparison window is anchored to the release marker. When multiple release markers fall within the comparison window, attribution is marked as ambiguous.

\subsection{Interestingness Scoring and Insight Prioritization}

We define a composite interestingness score adapted to the behavioral analytics setting:

\begin{equation}
  \text{score}(f) = \alpha \cdot \text{significance} + \beta \cdot \text{magnitude} + \gamma \cdot \text{reach} + \omega \cdot \text{actionability} + \varepsilon \cdot \text{novelty}
  \label{eq:score}
\end{equation}

where:
\begin{itemize}[leftmargin=1.5em]
  \item \textbf{significance} is the statistical confidence of the finding ($1 - p\text{-value}$), penalized for small sample sizes.
  \item \textbf{magnitude} is the normalized effect size: $|\text{lift} - 1|$ for conversion-related findings, $|\Delta P| / P_{t-1}$ for regression findings, normalized to $[0, 1]$.
  \item \textbf{reach} is the fraction of the user population affected by the finding. We distinguish this finding-level \emph{population reach} from the state-level \emph{reach rate} of Eq.~\eqref{eq:reachrate}, which is per-journey rather than per-user. The two coincide when each user contributes exactly one journey; in multi-journey-per-user products they diverge, and the scoring layer uses the per-user quantity to align with deployment impact.
  \item \textbf{actionability} is a heuristic score that favors findings for which product or engineering actions are known to exist.
  \item \textbf{novelty} measures the recency of the finding relative to prior snapshots.
\end{itemize}

The weights $(\alpha, \beta, \gamma, \omega, \varepsilon)$ are configurable per deployed monitoring context. We acknowledge that the definition of interestingness is inherently subjective \cite{silberschatz1995}.

Table~\ref{tab:scoring} illustrates the scoring algorithm applied to five synthetic findings from a representative product funnel. The prioritisation is well-behaved: the high-significance, high-reach activation driver (F002) ranks first; the statistically weak, small-sample drop-off cluster (F005) ranks last.

\begin{table}[htbp]
\centering
\caption{Composite interestingness scores for five synthetic detector findings
(weights: $\alpha{=}0.30$, $\beta{=}0.25$, $\gamma{=}0.20$, $\omega{=}0.15$,
$\varepsilon{=}0.10$). Scores computed from synthetic simulation data by
\texttt{interestingness\_scoring.py}.}
\label{tab:scoring}
\small
\begin{tabularx}{\linewidth}{cll>{\raggedright\arraybackslash}Xr}
\toprule
\textbf{Rank} & \textbf{ID} & \textbf{Detector} & \textbf{Finding (summary)} & \textbf{Score} \\
\midrule
1 & F002 & ActivationDriver   & \texttt{feature\_used} is activation driver for \texttt{converted} (lift 4.2$\times$) & 0.845 \\
2 & F001 & TemporalRegression & \texttt{email\_verified}$\to$\texttt{profile\_complete} drop-off $+$18\,pp post-v2.3  & 0.736 \\
3 & F004 & SegmentDivergence  & Mobile conversion 22\,pp below desktop                                               & 0.602 \\
4 & F003 & UnexpectedLoop     & Loop: \texttt{profile\_complete}$\to$\texttt{sign\_up} (8\% of users)                & 0.252 \\
5 & F005 & DropOffCluster     & Drop-off at \texttt{sign\_up}$\to$abandoned (low significance, $N{=}85$)             & 0.054 \\
\bottomrule
\end{tabularx}
\end{table}

\subsection{Confidence Scoring}

Each behavioral fact carries a confidence score in $[0, 1]$ computed as:

\begin{equation}
  \text{confidence} = \sigma\!\left(a \cdot z_{\text{score}} + b \cdot \log\frac{n}{n_{\min}} + c \cdot |\text{effect}|\right)
  \label{eq:confidence}
\end{equation}

where $\sigma$ is the logistic function, $z_{\text{score}}$ is the z-statistic, $n$ is the sample size, $n_{\min}$ is the configured minimum threshold, and $(a, b, c)$ are deployment-specific coefficients to be calibrated against ground-truth annotations once a deployed BIP instance accumulates them. Eq.~\eqref{eq:confidence} should therefore be read as defining a \emph{family} of confidence functions parameterised by $(a, b, c)$, rather than a closed-form result; the present paper does not commit to specific coefficient values, and the public simulation scripts do not implement Eq.~\eqref{eq:confidence}. Once calibrated, confidence is presented to end users as a qualitative label (High, Medium, Low).

\section{Detector System}
\label{sec:detectors}

\subsection{Design Principles}

The detector system operationalizes the claim that behavioral intelligence is generated, not queried. Each detector implements a specific detection hypothesis---a pattern in the behavioral graph that, if present and statistically supported, corresponds to a finding of a defined type. Detectors are deterministic: given the same snapshot inputs, they always produce the same outputs.

Detectors are versioned. When detector logic changes, previously produced findings are not modified; future snapshots are processed with the new version.

\subsection{Taxonomy of Detectors}

\subsubsection{Activation Driver Detector}

The activation driver detector identifies states strongly associated with conversion to a target terminal state. It: (1) computes reach rate and lift for all states, (2) filters to candidate states with $\text{reach\_rate} \geq \tau_{\text{reach}}$ and $\text{lift} \geq \tau_{\text{lift}}$ and $\text{sample\_size} \geq \tau_n$, (3) computes removal effects for candidate states, (4) ranks candidates by removal effect, (5) emits findings with predicate \texttt{is\_activation\_driver\_for} for top-ranked candidates.

\subsubsection{Drop-Off Detector}

The drop-off detector identifies states and edges with disproportionately high rates of journey termination without conversion. States with $\text{exit\_probability} \geq \tau_{\text{exit}}$ and $\text{reach\_rate} \geq \tau_{\text{reach}}$ are emitted as findings with predicate \texttt{is\_dropoff\_point\_for}. Clusters of consecutive drop-off states are aggregated into a single \texttt{dropoff\_cluster} finding.

\subsubsection{Behavioral Regression Detector}

The regression detector compares current snapshot metrics to a baseline and identifies statistically significant negative changes using the temporal change detection framework from Section~\ref{sec:formal}. Findings are emitted with predicates \texttt{regressed\_after} (for release-linked changes) or \texttt{changed\_after} (for unlinked changes).

\subsubsection{Segment Divergence Detector}

The segment divergence detector compares behavioral graph metrics across two or more user segments and identifies states or paths that exhibit significantly different conversion rates, reach rates, or transition probabilities. For segment pair $(S_1, S_2)$, it computes the Jensen-Shannon divergence of the transition probability distributions, a symmetric variant that does not require a canonical ordering of the two segments.

Findings are emitted with predicates \texttt{more\_common\_in} and \texttt{less\_common\_in} for reach rate differences, and \texttt{diverges\_from} for overall distribution divergence.

\subsubsection{Repeated Visit Detector}

The repeated visit detector identifies states visited multiple times within a single journey, indicating user confusion, friction, or a broken flow. A state $s$ is flagged when the mean visit count among journeys containing $s$ exceeds a threshold $\tau_{\text{loop}}$. Note that this detects unexpectedly high re-visitation frequency; detection of full cyclic paths (graph cycles) requires path-level tracking and is left to future work.

\subsubsection{Path Quality Detector}

The path quality detector evaluates materialized paths on a composite quality score:

\begin{equation}
  Q_{\text{path}} = \text{conversion\_rate} \times \frac{1}{\text{avg\_duration}} \times \frac{1}{\text{path\_length}}
  \label{eq:pathquality}
\end{equation}

High-quality paths are surfaced as findings with predicate \texttt{is\_fast\_path\_to}. Low-quality paths with high occurrence rates are flagged as candidate optimization targets. Note that average duration and path length are positively correlated; the multiplicative formulation jointly penalizes both dimensions. This is intentional for the present setting, where long and slow paths are doubly undesirable; however, practitioners may elect to drop one factor if the correlation is strong in a given dataset.

$Q_{\text{path}}$ is dimensionally heterogeneous (probability $\times$ inverse-time $\times$ inverse-length) and is therefore intended as a \emph{relative ranking} score within a single deployment, not as an absolute quantity comparable across deployments. Cross-deployment comparison requires either (i) per-deployment normalisation of duration and length to dimensionless quantities (e.g., dividing by the median path duration and length within the deployment), or (ii) replacing the inverse factors with bounded sigmoidal transforms over chosen reference scales.

\subsection{Detector Composition and Ordering}

Detectors run sequentially after each snapshot build. The orchestrator manages dependencies via a directed acyclic graph of detector jobs. Within each job, detectors are independently parallelizable.

\section{Grounded Language Layer}
\label{sec:gll}

\subsection{Architecture}

The GLL translates structured BKG findings and facts into natural language insights. The key architectural constraint is that the LLM must not perform any numerical computation, must not assert any claim not present in the fact bundle, and must not use causal language where only association is supported.

The GLL implements a three-stage pipeline: (i) fact bundle construction, (ii) grounding validation, and (iii) narrative generation.

\subsection{Fact Bundle Construction}

For a given finding $F$, the fact bundle $B(F)$ is constructed by retrieving from the BKG all facts that: (a) have \texttt{relatedEntityIds} intersecting the finding's entity set, (b) are valid within the finding's time window, and (c) exceed the minimum confidence threshold. Bundle construction uses a bounded-depth graph traversal (one-hop in the current implementation). This produces a structured context document such as:

\begin{minipage}{\linewidth}
\begin{lstlisting}
Finding: activation_driver
State: "import_data"
Predicate: is_activation_driver_for
Object: outcome:converted
Evidence:
  - reach_rate: 0.25
  - P(converted | reached): 0.58
  - P(converted | not reached): 0.03
  - lift: 19.88
  - removal_effect: 0.14
  - sample_size: 4,201
  - confidence: High
Supporting facts:
  - "import_data" transitions_to "invite_teammate" (p=0.70)
  - "import_data" is_dropoff_point_for "dropped_off" (p_dropoff=0.22)
\end{lstlisting}
\end{minipage}

\subsection{Grounding Validation}

Before generation, a grounding validator checks that all numerical values in the fact bundle are self-consistent (e.g., $P(\text{converted} \mid \text{reached})$ and $P(\text{converted} \mid \neg\,\text{reached})$ are within the valid probability range, lift is consistent with both probabilities, sample size meets the minimum threshold). Bundles failing validation are not passed to the LLM.

\subsection{Narrative Generation}

The validated fact bundle is passed to the LLM with a system prompt that enforces the following constraints: (i) report all numerical values exactly as provided in the fact bundle, (ii) use associative language (\textit{associated with}, \textit{more likely to}, \textit{suggests}) rather than causal language (\textit{causes}, \textit{drives}, \textit{leads to}), (iii) mention confidence and sample size where material, (iv) conclude with a concrete, actionable recommendation.

The resulting narrative is stored alongside the finding in the insight feed, with a provenance pointer back to the fact bundle and the LLM generation parameters.

\subsection{Agent Query Interface}

In addition to push-based insight delivery, the GLL supports a pull-based query interface. The query is processed by a structured planner that: (1) identifies the relevant journey definition and time window, (2) retrieves candidate facts from the BKG using semantic search over fact summaries, (3) constructs a bounded fact bundle from the retrieved facts, (4) generates a response constrained to the fact bundle. This architecture is analogous to GraphRAG applied to a dynamically constructed behavioral knowledge graph.

\section{Evaluation Framework and Open Measurement Challenges}
\label{sec:evaluation}

\subsection{Challenges Unique to Behavioral Intelligence Systems}

Evaluating behavioral intelligence systems presents challenges not present in standard supervised learning settings. There is no ground truth for the set of insights that should be generated from a given event stream. Evaluation must operate across multiple dimensions: computational correctness, statistical validity, and insight quality.

The Markov chain computations underlying Figures~\ref{fig:markov}, \ref{fig:absorption}, and~\ref{fig:removal}, as well as Table~\ref{tab:scoring} and the fact bundle in Section~\ref{sec:gll}, are reproduced exactly by the public simulation scripts at \url{https://github.com/journium/journium-research}: closed-form computation in \texttt{markov\_journey\_model.py} and \texttt{removal\_effect.py}, Monte Carlo verification of empirical conditional conversion rates in \texttt{simulate\_trajectories.py}, and composite scoring in \texttt{interestingness\_scoring.py}. All figures serve as worked examples of formal correctness, not as empirical evaluation of a deployed system.

\subsection{Correctness Metrics}

Computational correctness is evaluated through: (i) deterministic consistency---for identical input snapshots, the system must produce identical outputs; (ii) traceability---for a random sample of facts, numerical values in the evidence payload must match recomputed values from the raw snapshot data. A target of $\geq\!99\%$ traceability is achievable with deterministic computation.

For the GLL, faithfulness is evaluated by extracting verifiable claims from the narrative and checking each against the fact bundle, analogous to the claim-verification methodology of InsightBench \cite{sahu2024}.

\subsection{Insight Quality Metrics}

Insight quality is evaluated along three dimensions: precision (are surfaced insights genuinely interesting?), recall (are important behavioral phenomena detected?), and actionability (do insights lead to product actions?). Initial evaluation targets are: activation driver detector precision $\geq 70\%$, regression detector false positive rate $\leq 15\%$, and narrative summaries rated `actionable' by $\geq 75\%$ of evaluators.

\subsection{Latency and Throughput}

Core BGE snapshot builds should complete within 1 hour for projects with up to 10M events per day. Detector pipeline execution should complete within 24 hours of the analysis window closing. GLL narrative generation should complete within seconds per finding.

\subsection{Comparison with Pull-Based Baselines}

BIP is compared against a pull-based baseline on a set of evaluation tasks: identifying the top activation driver, identifying the most significant behavioral regression in a post-release period, and identifying the primary segment divergence. We measure detection rate, time-to-insight, and precision.

\section{Discussion}
\label{sec:discussion}

\subsection{Design Principles}

BIP is built on four design principles that merit explicit discussion.

\textbf{Grounded-first computation.} Every insight in the feed must be traceable to a deterministic computation. The LLM is a linguistic interface, not a reasoning engine. The cost of hallucinated behavioral insights in a production analytics tool is high: misallocated engineering effort, incorrect product decisions, and eroded trust. Grounding is non-negotiable.

\textbf{Semantic abstraction hierarchy.} The three-level state model (raw, semantic, lifecycle) balances expressiveness and tractability. Without semantic abstraction, the graph has too many nodes for meaningful insights; with too much abstraction, the graph loses the specificity needed to guide product decisions.

\textbf{Time-aware by default.} All metrics, facts, and insights are snapshot-scoped. This enables temporal comparison, release-linked change detection, and trend analysis. The decision to make snapshots immutable prioritizes auditability over storage efficiency.

\textbf{AI on top of truth.} The GLL architecture enforces that language generation is strictly downstream of verified computation. The fact schema must carry sufficient numerical detail to enable full regeneration of a faithful narrative, without the LLM needing to infer or compute \cite{pan2024}.

\subsection{Limitations and Open Problems}

\begin{itemize}[leftmargin=1.5em]
  \item \textbf{Causal inference.} Removal effects and lift scores are not causal effects. Identifying which behavioral patterns causally drive activation or conversion requires controlled experimentation or instrumental variable methods beyond the scope of the current architecture.

  \item \textbf{Non-stationarity.} The Markov assumption and the stationarity assumption are known to be violated in practice. Addressing non-stationarity through time-varying Markov models or mixture models \cite{harbich2017} is a priority for future work.

  \item \textbf{Cross-journey dependencies.} The journey model treats each journey instance as independent, ignoring network effects, referral chains, and account-level interactions.

  \item \textbf{Learned representations.} The current architecture uses shallow, hand-crafted features. Deep learned representations of event sequences could capture higher-order patterns not visible to the Markov model.

  \item \textbf{Evaluation of interestingness.} The interestingness score defined in Section~\ref{sec:formal} is a principled approximation, but there is no ground truth for what makes an insight interesting. Developing automated proxies that correlate with human judgment is an open research problem.

  \item \textbf{Privacy and data minimization.} State definitions must be constrained to operate on an allowlist of safe properties, and evidence payloads must not store raw user attributes. Differential privacy techniques for behavioral graph statistics are an important direction.
\end{itemize}

\subsection{Broader Impact}

By automating the generation of behavioral insights, BIP has the potential to democratize product intelligence---making sophisticated behavioral analysis accessible to product teams that lack dedicated data science resources. The system is designed to assist and augment human judgment, not replace it; all insights include the evidence and confidence information needed for informed human decision-making.

\section{Conclusion}
\label{sec:conclusion}

We have presented the Behavioral Intelligence Platform (BIP), a formal architecture for transforming raw product event streams into automatically generated, evidence-backed narrative insights. BIP introduces four tightly integrated layers: a Normalization and State Derivation stage that standardizes raw events and derives a multi-level semantic state hierarchy; a Behavioral Graph Engine that models user journeys as absorbing Markov chains; a Behavioral Knowledge Graph and Detector System that reify graph outputs into a typed triple-store of grounded facts and autonomously identify behavioral phenomena; and a Grounded Language Layer that constrains LLM-generated narratives to verified facts from the BKG.

The architecture is grounded in a formal treatment of the Behavioral Intelligence Problem, with a well-defined taxonomy of detectors, a principled interestingness scoring framework for insight prioritization, and a confidence model that enables calibrated communication of statistical uncertainty. The core design principles---grounded-first computation, semantic abstraction, time-awareness, and AI on top of truth---reflect lessons from adjacent fields including process mining, automated data exploration, knowledge graphs, and retrieval-augmented generation.

BIP represents a step toward analytics systems that take epistemic initiative: systems that do not wait to be asked, but continuously monitor, detect, and narrate the behavioral landscape of a product. The fundamental challenges of causality, non-stationarity, and evaluation of interestingness remain open, and we look forward to collaborative research progress on these problems.


\bibliographystyle{plain}
\bibliography{bibliography}

\end{document}